\documentclass[preprint,showpacs,preprintnumbers,amsmath,amssymb]{revtex4}
\usepackage{graphicx}
\pacs{}
\begin{document}
\def\v#1{\mathbf{#1}}
\def\vt{\tilde{V}}
\def\half{\frac{1}{2}}
\def\d#1{#1^{\dagger}}
\def\om{\Omega}
\def\omr{\omega}
\def\tomr{\tilde{\omr}}
\def\ep{\epsilon}
\title{Anomalous transport and phonon renormalization in a chain with transverse and longitudinal vibrations}
\author{Santhosh G}
\affiliation{The Institute of Mathematical Sciences, C.I.T. Campus, Chennai 600 113, India. }
\author{Deepak Kumar}
\affiliation{Jawaharlal Nehru University, New Delhi 110 067, India.}

\begin{abstract}
We study thermal transport in a chain of coupled atoms, which can vibrate in longitudinal as well as transverse directions. The particles interact through anharmonic potentials upto cubic order. The problem is treated quantum mechanically. We first calculate the phonon frequencies self-consistently taking into account the anharmonic interactions. We show that for all the modes, frequencies must have linear dispersion with wave-vector $q$ for small $q$ irrespective of their bare dispersions. We then calculate the phonon relaxation rates $\Gamma_i(q)$, where $i$ is the polarization index of the mode, in a self-consistent approximation based on second order perturbation diagrams. We find that the relaxation rate for the longitudinal phonon, $\Gamma_x(q) \propto q^{3/2}$, while that for the transverse phonon $\Gamma_y(q) \propto q^2$. The consequence of these results on the thermal conductivity $\kappa(N)$ of a chain of $N$ particles is that $\kappa(N) \propto N^{1/2}$. 
\end{abstract}
\pacs{44.10.+i,05.70.Ln,05.60.Gg,66.25.+g}
\maketitle

\section{Introduction}
In low dimensional systems, the description of transport in terms of linear phenomenological laws, that works well in three dimensions, may not be possible. One example is the Fourier law which relates the thermal current density $j$ to the thermal gradient $\nabla T$  by $j=-\kappa \nabla T$, where the $\kappa$ is the thermal conductivity of the material. Theoretical investigations of a number of translation invariant one-dimensional systems show that the Fourier law  is not followed in these systems \cite{Bonetto}. The ratio of the current density to the thermal gradient is no longer independent of the system size. Nevertheless this ratio, again called `thermal conductivity', is still useful in characterizing anomalous transport.

The calculation of `thermal conductivity' in these systems can be problematic. Ideally one would like to calculate the non-equilibrium stationary state (NSS) measure of the system and calculate the expectation values of $j$ and $\nabla T$ using the NSS measure. This requires the existence of local thermal equilibrium (LTE) so that a local temperature $T$ is properly defined. A method generally used when considering a one dimensional system with its ends at two different temperatures $T_1$ and $T_2$, is to calculate directly the current $j$ and the quantity $\kappa(L)=jL/(T_2-T_1)$, $L$ being the length of the system. Another approach is to use the Kubo-Green formula which involves the time integral of the equilibrium current-current correlation function. The first formula does not require the existence of LTE whereas the second does and it is not clear how the results from the two methods can be compared. 
 
In 1-D systems with translation invariance, the `conductivity', $\kappa$, shows a power-law dependence on the size of the system; $\kappa \sim L^{\alpha}$~\cite{LepriLiviPolitiPhysRep}. A question that has been investigated extensively is whether $\alpha$ is universal across a class of systems. The other related questions are about the number of such universality classes and the values of $\alpha$ characterizing them. Such an investigation has to take into account the difficulties posed by the considerations in the previous paragraph.  Hydrodynamic mode-coupling theory predicts two universality classes corresponding to $\alpha=1/2$ and $\alpha=1/3$ for one-dimensional momentum conserving systems~\cite{LeeNickelGrayPRE2005,NarayanRamaswamy}. Specific models of one-dimensional lattice systems with pair-wise nonlinear interactions between neighboring particles have been studied extensively using numerical and analytical methods~\cite{LepriLiviPolitiPhysRep}. The analytical studies involve approximations at some stage, whereas numerical studies are handicapped by the system size and the time scales that can be achieved in computers which makes it difficult to give a definitive answer to the question of the universality of $\alpha$. Results of these studies agree with the hydrodynamic results that the generic value of $\alpha$ is $1/3$; when the leading order nonlinearity is cubic $\alpha=1/3$~\cite{SanthoshDeepakOddPot}. However, for the so-called Fermi-Pasta-Ulam chain, where the nonlinearity is quartic, there is still no consensus; numerical simulations show different values of $\alpha$~\cite{llp98,mdn} and analytic calculations, both classical and quantum, show $\alpha=2/5$~\cite{Perverzev,Lepri,SanthoshDeepakFPUbeta}.

Apart from the obvious theoretical interest in these problems there has also been a lot of experimental interest in studying heat transport at micro and nano scales \cite{Volz}. Recent experimental studies on nanowires such as Carbon nanotubes and Boron-nitride nanotubes show a clear breakdown of the Fourier law~\cite{ChangExptPRL,Kim,Hone,Chang}. In order to attempt an understanding of realistic systems, it is essential to include all the modes of vibrations. Theoretical results, both numerical and analytical, mentioned above have been obtained for only the longitudinal vibrations of the chain. It is natural to investigate how the exponent $\alpha$ is affected by the transverse vibrations.  

Chains with both longitudinal and transverse vibrations have only been studied in the work of Wang and Li ~\cite{WangLi}. The Wang-Li model contains two kinds of intractions; a pairwise interaction between the neighboring particles and another interaction that depends on the bending of the chain. The harmonic approximation of the model yields modes whose frequencies $\Omega_{\alpha}(q)$ depend on the wave-vector $q$ in the following manner. For small q, the longitudinal modes have $\Omega_x(q)\propto q $ whereas the transverse modes $\Omega_y(q) \propto q^2$. Wang and Li used the mode-coupling analysis (MCA) and molecular dynamics (MD) simulations to study this system. Their first important observation from MD simulations is that the phonon dispersions are strongly renormalized from their bare values, particularly the transverse phonon dispersion gets renormalized to become linear in q. For this reason they use an effective Hamiltonian method in which the form of the Hamiltonian and its parameters are taken from the MD simulations. For the conductivity exponent they obtain a generic value $\alpha=1/3$, but they also observe a value of $\alpha=2/5$ in their MD simulations which is attributed to crossover effects. When the parameter measuring the strength of transverse vibrations is sufficiently large, they also find that $\alpha=1/2$. 

In this paper, we study the quantum mechanical version of this model along the lines of ~\cite{SanthoshDeepakOddPot, SanthoshDeepakFPUbeta}. In this procedure, one essentially calculates the second order self-energy of the phonons self-consistently, which enables us to obtain the wave-vector dependent relaxation rate $\Gamma (q)$. The thermal conductivity of the chain is then obtained through the use of Kubo formula. The finite size of the chain is taken into account by putting on the time integral in the Kubo formula an upper cutoff proportional to the size of the chain \cite{LepriLiviPolitiPhysRep}. This prescription yields $\alpha = 1- 1/\delta$, where $\delta$ governs the small-q behavior of the relaxation rate, $\Gamma (q) \propto q^{\delta}$. For the Wang-Li model, we find that in the second-order perturbation calculation, the relaxation rate $\Gamma_y(q)$ for the transverse phonon diverges as $q \rightarrow 0$, whereas the relaxation rate $\Gamma_x(q)$ for the longitudinal phonons becomes constant. This makes the bare phonons invalid normal modes. These divergences basically arise due to the $q^2$-dispersion of the transverse modes. This necessitates a self-consistent study of the phonon dispersion for this system \cite{Maradudin,Gotze,Horner,Horton}. Such studies have been done on several systems, including one-dimensional systems as test cases for the procedure \cite{Cuccoli,Freidkin,Schirmacher}. Here we follow the method of G\"{o}tze and Michel \cite{Gotze} which provides a nice connection between the Mori-Zwanzig mode coupling technique and the perturbation theory. We show that this procedure indeed renormalizes self-consistently the transverse phonon dispersion to be linear, in qualitative agreement with the numerical results of MD simulations \cite{WangLi}. We further note that the result on the renormalized frequencies is independent of the bare frequencies, but does depend on the fact that the cubic interaction vanishes linearly with q. Since the vanishing of the interaction with any of its wave-vector arguments is a consequence of the translation invariance, this result should be generic to translation invariant one-dimensional systems.
 
We then perform a self-consistent analysis for the relaxation rates of the renormalized modes using the second order perturbation diagrams. This procedure leads to integral equations from which small-q dependence of the relaxation rates $\Gamma_{\alpha}(q)$ can be extracted. The main results of this analysis are: $\Gamma_x(q) \propto q^{3/2}$ and $\Gamma_y(q) \propto q^2$. Gratifyingly, this behavior matches the results derived by Wang and Li for the classical system using the mode-coupling analysis, which arrives at the  results through a seemingly different mechanism. Finally we use these results for understanding the behavior of conductivity with the system length $L$. 

This paper is organized in the following manner. In Sec. II, we present the model and a calculation of the relaxation rates of the bare phonons. As mentioned above, this calculation shows that the modes are not well defined. Accordingly in Sec. III, we present a self-consistent calculation for the frequencies of the modes, which shows that both the modes should disperse linearly at small wave-vectors. In Sec. IV, we calculate the relaxation rates of the renormalized phonons using second order perturbation theory. In section V,  we formulate equations for the relaxation rates using the self-consistent second order perturbation theory and obtain their behavior at small wave-vectors.  In Sec. VI, we turn to the calculation of thermal conductivity using the finite size procedure mentioned above. We close the paper with a summary and discussion of our results in Sec. VII.
     
\section{Wang and Li's Model}
This model consists of a chain of equal mass point particles with two kinds of interactions. First is a pairwise interaction between nearest-neighbor particles which depends only on the absolute value of the distance between the particles and the second is a three particle interaction between neighbors that takes into account the bending of the chain. There are two transverse directions, but due to the axial symmetry of the chain the two directions are entirely equivalent. Accordingly for understanding the qualitative role of the transverse vibrations on the relaxation of modes and transport, it suffices to consider just one transverse mode. In terms of the position $\v{r}\equiv(x,y)$ and momentum $\v{p}=(p_x,p_y)$ vectors, the Hamiltonian is written as
\begin{equation} 
\label{HamiltonianFull}
 H=\sum_{l} \Big [ \frac{\v{p}^2(l)}{2} +\half K_r \left( |\v{r}(l+1)-\v{r}(l)|-a\right)^2
+ K_{\theta} \cos(\theta(l))
\Big ],
\end{equation}
where $a$ is the lattice constant, $\cos(\theta(l))=-\v{n}(l-1)\cdot \v{n}(l)$ and $\v{n}(l)$ is the unit vector along $\v{r}(l+1)-\v{r}(l)$. The mass of the particles has been set to unity. This is a complicated many body problem. One assumes that a few lower-order terms in the Taylor expansion of the nonlinear potential is enough to give the generic physical picture. Also the low-temperature transport properties of the system are expected to be described by the low-energy modes of the system. So we set up the problem in terms of phonon modes. Denoting the deviation from equilibrium position of the particles by $\v{\phi}$ we have $\v{r}(l)=(l a+\phi_x(l), \phi_y(l))$ and $\v{p}(l)=(\pi_x(l),\pi_y(l))$.  Fourier transform along the length of the chain is defined by
\begin{equation}
f(k)= \frac{1}{\sqrt{N}} \sum_{l=1}^N e^{-i2\pi k l} f(l),
\end{equation}
where $N$ is the number of particles in the chain. The Hamiltonian, considering nonlinearities up to cubic terms only, can be written as~\cite{WangLi}
\begin{eqnarray}\label{HamiltonianCubic}
 H=\frac{1}{2}\sum_{k i}\pi^{\dagger}_i(k)\pi_i(k)+\frac{1}{2!}\sum_{\v{k}\v{i}}V_2(k_1i_1,k_2i_2)\phi^{\dagger}_{i_1}(k_1) \phi^{\dagger}_{i_2}(k_2)\nonumber \\
+\frac{1}{3!\sqrt{N}}\sum_{\v{k}\v{i}}V_3(k_1i_1,k_2i_2,k_3i_3) \phi^{\dagger}_{i_1}(k_1) \phi^{\dagger}_{i_2}(k_2)\phi^{\dagger}_{i_3}(k_3),
\end{eqnarray}
where the summation indices in bold letters indicate that the summation is over all the polarization indices $i_1, i_2,..$ which takes values from $\{x,y\}$ and over all the wave-vectors $k_1, k_2, ..$, a notation we will follow hereafter. $V$'s are symmetric under exchange of pairs $k_m i_m \leftrightarrow k_n i_n$. Translation invariance  imposes two conditions on $V$; they vanish when any of the wave-vectors is set to zero and $V_n(k_1 i_1,.., k_n i_n) \propto \Delta(k_1+..+k_n) $, where $\Delta(k)$ is unity when $k$ is a reciprocal lattice vector and zero otherwise. Explicit forms for the $V$'s are obtained by expanding the potential term in Eq.~(\ref{HamiltonianFull}) in Taylor series and taking the Fourier transforms. The harmonic terms are given by~\cite{WangLi}
\begin{eqnarray}
 V_2(k_1 i_1,k_2 i_2)=\delta_{k_1+k_2} \delta_{i_1,i_2} \om_{i_1}^ 2(k_1),\nonumber \\
\om_x(k)=2 \om_x |\sin(ka/2)|,\quad \om_y(k)=4 \om_y \sin^2(ka/2), \nonumber\\
\om_x^2=K_r, \quad \om_y^2=K_{\theta}/a^2.
\end{eqnarray}
Note that the frequency of the longitudinal mode vanishes linearly with the wave-vector whereas for the transverse mode it vanishes quadratically. Henceforth we choose units in which $a=1$. The reflection symmetry in the transverse direction permits only two kinds of cubic terms to be nonzero, these being $V_3(k_1 x,k_2 y,k_3 y)$ and $V_3(k_1 x,k_2 x,k_3 x)$.  Both these interactions are proportional to $\sin(k_1/2) \sin(k_2/2) \sin(k_3/3)$. These forms are generic when the potentials are taken to be functions of absolute values of relative displacements ($|\v{r}_{i+1}-\v{r}_i|$)\cite{WangLi}.  

For the Fermi-Pasta-Ulam chain~\cite{SanthoshDeepakFPUbeta} in which only the longitudinal vibrations are considered, the perturbation analysis is done around the bare phonon modes as they are well defined. But in this model we shall see that the bare phonon modes are not well defined, as the relaxation rates of the two kinds of phonons are not zero even for zero wave-vector mode. This requires that one works with the renormalized phonons as discussed below. We begin by calculating the relaxation rates of the bare phonons. For this purpose we first define the bare phonon operators
\begin{equation} \label{aDefinition}
 a_{i}(k)=\sqrt{\frac{\om_i(k)}{2}} \left( \phi_i(k)+i \frac{\pi_i(k)}{\om_i(k)} \right),
\end{equation}
in terms of which the quadratic part of the Hamiltonian becomes diagonal. The Hamiltonian can be written as
\begin{eqnarray}\label{HcubicInNumberOp}
 H=\sum_{k i}\om_i(k) \d{a}_i(k)a_i(k)+\frac{1}{3!\sqrt{N}}\sum_{\v{k}\v{i}} \vt_3(k_1 i_1, k_2 i_2, k_3 i_3) U_{i_1}(k_1) U_{i_2}(k_2) U_{i_3}(k_3),
\end{eqnarray}
where  we have defined $U_i(k)=a_i(k)+\d{a}_i(-k)$ and $\vt_3(k_1 i_1, k_2 i_2, k_3 i_3)=V_3(k_1 i_1, k_2 i_2, k_3 i_3)/\sqrt{\om_{i_1}(k_1)\om_{i_2}(k_2)\om_{i_3}(k_3)}$.  
\begin{figure}[t]
\begin{center}
\includegraphics  [width=15cm]{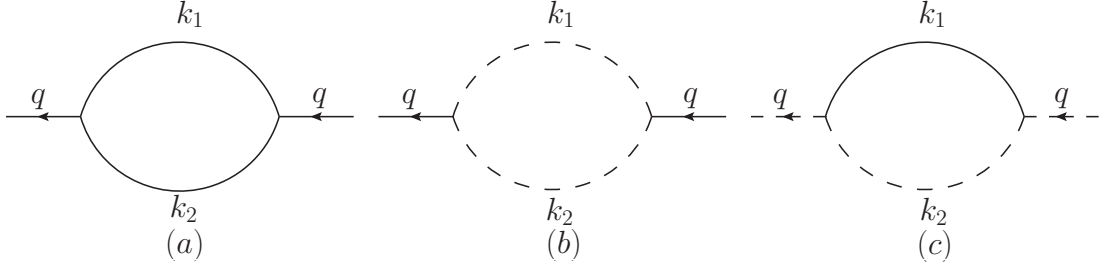}
\caption{Diagrams for the second order contribution to the self-energy $\Sigma^{(2)}_{j}(q)$. 
The bare phonon Green's functions $D^{(0)}_x(k)$ for the longitudinal phonons and $D^{(0)}_y(k)$ for the transverse phonons are represented respectively by the solid lines and the dashed lines. The diagrams (a) and (b) are two contributions for the longitudinal phonons, while the diagram (c) gives the sole contribution for the transverse phonons.}
\label{selfEnergyDiagrams}
\end{center}
\end{figure}

For a system with two phonon polarizations, in general one defines the single particle Green's functions as the time-ordered expectation values 
$ G_{i,j}(q,\tau)=-\left \langle T_{\tau}\left[ a_i(q,\tau) \d{a}_j(q,0) \right ] \right\rangle$, where the time dependence of operators is given by $O(\tau)=e^{\tau H}Oe^{-\tau H}$, $\tau$ being the imaginary time. The angular brackets correspond to thermal averages defined by $\left \langle O \right \rangle=Tr[e^{-\beta H} O]/Tr[e^{-\beta H}]$. The $G$'s are thus 2x2 matrices in polarization indices, but in the present case, the reflection symmetry in the transverse direction allows only the diagonal components to be nonzero. In the following we just write them as $ G_i(q,\tau)$. The perturbation series for $G_j$ can be summed into the form $G_j^{-1}(q,i\omega_n)=i\omega_n-\om_j(q)-\Sigma_j(q,i\omega_n)$, where $\omega_n$'s denote the usual Matsubara frequencies and $\Sigma_j(q,i\omega_n)$ is the self-energy. The relaxation rate is defined as $\Gamma_j(q,\omega)=-\Im \Sigma_j(q,i\omega_n \to \omega+i0^+)$, and the on-shell relaxation rate is denoted by $\Gamma_j(q)\equiv \Gamma_j(q,\om_j(q))$. We first evaluate the relaxation rate up to second-order in the perturbation series. The second-order diagrams for $\Sigma_x$ that contribute to the relaxation rate are given in Figs.~\ref{selfEnergyDiagrams} (a) and (b). Their contribution is
\begin{eqnarray}\label{sigma}
 \Sigma^{(2)}_x(q,\tau)=-\frac{1}{2N} \sum_{k_1,k_2,i} |\vt_3(-q x,k_1 i,k_2 i)|^2 D_i^{(0)}(k_1,\tau) D_i^{(0)}(k_2,\tau),
\end{eqnarray}
where $D_i^{(0)}(q,\tau)$ is the following unperturbed Green's function.  
\begin{equation}
D_i^{(0)}(q,\tau)=G_i^{(0)}(q,\tau)+G_i^{(0)}(-q,-\tau)=-\sum_{s\in\{+,-\}} e^{-s\om_i(q)\tau} n_s(\om_i(q)),
\end{equation}
where $n_-(\om)=n(\om)$ is the Bose factor and $n_+(\om)=1+n(\om)$. The relaxation rate is given by 
\begin{eqnarray}
 \Gamma_x^{(2)}(q)&=&\frac{\pi}{2N}(1-e^{-\beta \om_x(q)}) \sum_{k_1,k_2,s_1,s_2,i}  
|\vt_3(-q x,k_1 i,k_2 i)|^2  n_{s_1}(\om_i(k_1)) n_{s_2}(\om_i(k_2)) \nonumber\\ 
&&\times \delta(\om_x(q)-s_1\om_i(k_1)-s_2\om_i(k_2)).
\end{eqnarray}
The sum is simplified by two conditions: the $\Delta$-function contained in the potential $V_3$ and the $\delta$-function. Therefore, we can write
\begin{eqnarray}
 \Gamma_x^{(2)}(q)&=&\frac{1}{4} (1-e^{-\beta \om_x(q)})\int dk \sum_{s_1 s_2\,i} |\vt_3(-q\, x,k\, i,q-k\, i)|^2  n_{s_1}(\om_i(k)) n_{s_2}(\om_i(q-k)) \nonumber\\ 
&&\times \delta(\om_x(q)-s_1\om_i(k)-s_2\om_i(q-k))\nonumber \\
\label{gamm2BarePhonon}
&=&\frac{1}{4} (1-e^{-\beta \om_x(q)})\sum_{s_1,s_2,i,k^*}  \frac{|\vt_3(-q x,k^*\, i,q-k^*\, i)|^2}{|J_i(q,k^*,s_1,s_2)|}  n_{s_1}(\om_i(k^*)) n_{s_2}(\om_i(q-k^*)),
\end{eqnarray}
where $k^*\equiv k^*(q,s_1,s_2,i)$ is the solution of $f_i(q,k,s_1,s_2):=\om_x(q)-s_1\om_i(k)-s_2\om_i(q-k)=0$ and the Jacobian $J_i(q,k^*,s_1,s_2)=(\partial/\partial k)f_i(q,k,s_1,s_2)|_{k=k^*}$. If the sum on the RHS of Eq.~(\ref{gamm2BarePhonon}) is finite we expect $\Gamma_x^{(2)}(q\to 0) \propto \om_x^2(q)$, as $|\vt_3(-q\, x,k\, i,q-k\, i)|^2 \propto \om_x(q)$. We first note that the term corresponding to $i=x$ (Fig. 1(a)) which involves only the longitudinal modes gives zero contribution as the condition $f=0$ can have only trivial solutions at which the self-energy itself becomes zero ~\cite{SanthoshDeepakOddPot}. 

The second term corresponding to $i=y$ (Fig. 1(b)) needs to be examined for singular behavior. The source for singularities are the Jacobian $J$ and the number factors which for low frequencies have the form $n(\om\to 0) \sim (\beta \om)^{-1}$. For this term we have transverse mode frequencies which vanish quadratically with the wave-vector and hence are more singular. Considering $s_1=+=s_2$ term, the small-$q$ solution is obtained to be $k^*(q)\sim \sqrt{|q|}$ and the Jacobian $J(q,k^*)\sim \sqrt{|q|}$. This yields a contribution to $\Gamma_x(q)$ which diverges for small $q$ at finite temperatures and we have the leading behavior $\Gamma_x(q \to 0) \sim |q|^{-1/2}$.

The second-order diagram for $\Sigma_y$ is given in Fig. 1(c).  The yields the relaxation rate as
\begin{eqnarray}
 \Gamma_y(q)&=&\half \left( 1-e^{-\beta \om_y(q)} \right ) \int dk \sum_{s_1 s_2} |\vt(-q y,k x,q-k\, y)|^2  n_{s_1}(\om_x(k)) n_{s_2}(\om_y	(q-k)) \nonumber\\ 
&&\times \delta(\om_y(q)-s_1\om_x(k)-s_2\om_y(q-k)).
\end{eqnarray}
Proceeding as in the analysis of Eq.~(\ref{gamm2BarePhonon}), we see that the term corresponding to $s_1=+, s_2=-$ is finite for $q\to 0$. 
\section{Renormalization of phonons}
The above analysis shows that the bare phonon modes are not stable under scattering and thus are not good zeroth order approximations in the perturbation theory. Therefore we follow the mode-coupling method as formulated by G\"{o}tze and Michel~\cite{Gotze} to study how the bare phonon dispersions get affected by the nonlinearities. For completeness we first present a brief outline of the mode-coupling method~\cite{Gotze}. The procedure requires evaluation of scalar products defined in the space of the dynamical operators (denoted by $O_1,O_2,..$) as follows.
\begin{eqnarray}
(O_1,O_2)&=&-\ll \d{O}_1; O_2\gg_{z=0}, \nonumber\\
\label{chiDefinition}
\ll O_1; O_2\gg_{z}&=&\pm \int dt \Theta(\pm t) e^{izt} \left \langle \left [O_1(t),O_2(0) \right] \right \rangle, \quad\quad \Im z\gtrless0, 
\end{eqnarray}
where $O(t)= e^{i\mathcal{H} t} O \equiv e^{iHt} O e^{-iHt}$. For us the key quantity is the static susceptibility $D^{-1}_{ij}(q)=\left( \phi_i(q), \phi_j(q) \right )$. It is easily seen that in the harmonic approximation $D_{ij}(q) = \om^2_i(q) \delta_{ij}$. The renormalized frequencies are the eigenvalues of the matrix $\v{D}(q)$ for the nonlinear problem.  It follows from the definition and the space reflection symmetry of the Hamiltonian that $\v{D}(q)$ is a real symmetric matrix which is an even function of $q$. Therefore, $\v{D}$ is diagonalizable with a complete and orthonormal set of eigenvectors $e^{\alpha}$ obeying
\begin{equation}
\label{Renorm}
 \sum_j D_{ij}(q) e^{\alpha}_j(q) = \omr^2_{\alpha}(q) e^{\alpha}_i(q),
\end{equation}
where the eigenvalues obey $\omr_{\alpha}(q)=\omr_{\alpha}(-q)$ due to time-reversal symmetry. 

 In order to develop equations for these scalar products, we utilize the identity, $\left( O_1,\mathcal{H} O_2 \right)=\langle [\d{O}_1,O_2] \rangle$ which follows from Eq.~(\ref{chiDefinition}). The LHS of this equation can be calculated using $\mathcal{H} O=[H,O]$. For further calculations we define $ A_i(q)= \half\sum_{\v{k}\v{i}} V(q i,-k_1 i_1, -k_2 i_2) \phi_{i_1}(k_1) \phi_{i_2}(k_2)$. Taking (i) $O_1=\phi(q), O_2=\pi(q)$ and (ii) $O_1=A(q), O_2=\pi(q)$ respectively, we get the following equations which are exact.
\begin{eqnarray}
 D_{ij}(q)=V(-q i, qj)+ \sum_k \left( A_i(q),\phi_k(q) \right) D_{kj}(q), \\
\sum_k (A_i(q), \phi_k(q))V(-qk,ql)=-(A_i(q),A_l(q)).
\end{eqnarray}
We write these equations in the matrix form, suppressing the $q$-dependence, as
\begin{equation}\label{DEquation}
\v{D}= \v{V}[\v{V}+(\v{A},\v{A})]^{-1} \v{V},
\end{equation}
where $(\v{A},\v{A})_{ij}(q)= (A_i(q),A_j(q))=-\ll A_i(q);A_j(q) \gg_{z=0}$ and $\v{V}_{ij}(q)=V_2(-qi,qj)$. To get a closure for this set of equations, we need to make some approximations. We do this in a self-consistent manner by expressing the higher order correlation functions $(A_i(q),A_j(q))$ in terms of $\v{D}$ matrices. The expression for $(\v{A},\v{A})$ is is given by
\begin{eqnarray}\label{EqAA}
\ll A_i(q);A_j(q)\gg_z &=& \frac{1}{4} \sum_{\v{k}\v{i}} V_3(-qi,-k_1 i_1,-k_2 i_2) V_3(qj,-k_3 i_3,-k_4 i_4)\nonumber \\
&& \times \ll \phi_{i_1}(k_1) \phi_{i_2}(k_2); \phi_{i_3}(k_3) \phi_{i_4}(k_4)\gg_z.
\end{eqnarray}
The expectation contained in the above equation is evaluated using an effective quadratic Hamiltonian given by
\begin{eqnarray}
 H_0 &=& \half\sum_{k i} \d{\pi}_i(k) \pi_i(k) + \half\sum_{k \v{i}} \d{\phi}_{i_1}(k) D_{i_1 i_2}(k) \phi_{i_2}(k), \nonumber \\
     &=& \sum_{k,\alpha} \left[ \omr_{\alpha}(k) \d{b}_{\alpha}(k)b_{\alpha}(k)+\half \right],
\end{eqnarray}
where the operators $b_{\alpha}(k)$ are the bosonic operators defined in the usual manner as 
\begin{equation}
 b_{\alpha}(k)=\sum_{j} e^{\alpha}_j(k)  \left [ \sqrt{2 \omr_{\alpha}(k)} \phi_j(k)+i\sqrt{2/\omr_{\alpha}(k)} \pi_j(k) \right ].
\end{equation}
Here the frequencies $\omr_{\alpha}(k)$ and the polarization vector $e^{\alpha}_i(k)$ are defined in Eq. (\ref{Renorm}), and are to be determined self-consistently.

Under this approximation using Eq.(~\ref{chiDefinition}) we get 
\begin{eqnarray}\label{AAEqExpanded}
 (A_i(q),A_j(q))=2\sum_{1,2} \mathcal{G}_{ij}(q,1,2)  \left\{ \frac{1+n(1)+n(2)}{\omr(1)+\omr(2)} +\frac{n(1)-n(2)}{\omr(2)-\omr(1)} \right\} \nonumber \\
\mathcal{G}_{ij}(q,1,2)=\sum_{\v{k} \v{i}}\frac{V_3(-qi,-k_1 i_1,-k_2 i_2) V_3(qj,k_1 i_3,k_2 i_4)}{8\omr(1)\omr(2)} e^{\alpha_1}_{i_1}(k_1)e^{\alpha_2}_{i_2}(k_2)e^{\alpha_1}_{i_3}(k_1)e^{\alpha_2}_{i_4}(k_2),
\end{eqnarray}
where we have used the compact notation $1\equiv (k_1,\alpha_1)$. Further $\omr(1)=\omr_{\alpha_1}(k_1)$ and $n(1)=n(\omr(1))$. Eqs.~(\ref{DEquation}) and (\ref{AAEqExpanded}) together provide us the self-consistent equations for determining the renormalized frequencies $\omr_{\alpha}(q)$. In order to obtain their solution, we first iterate them. Starting with a diagonal  $D_{ij}(q)=\delta_{ij}\omr^2_i(q)$ in the RHS of Eq.(~\ref{AAEqExpanded}), we have $e^{\alpha}_i=\delta_{i \alpha}$ and we see that only the diagonal terms survive. So this procedure gives a diagonal $\v{D}$ back if our starting $\v{D}$ is diagonal. Putting this information in Eq.~(\ref{DEquation}) reduces it to
\begin{eqnarray}\label{DEqDiagonal}
D_{\alpha \alpha}(q)&=&\frac{\omr_{\alpha}^4(q)}{\omr_{\alpha}^2(q)+(A_{\alpha}(q),A_{\alpha}(q))},\nonumber \\
(A,A)_{xx}(q)&=&\frac{1}{4} \sum_{\v{k} i} \frac{|V_3(qx,k_1 i,k_2 i)|^2}{\omr_{i}(k_1)\omr_{i}(k_2)} \left \{  \frac{1+n(\omr_i(k_1))+n(\omr_i(k_2))}{\omr_i(k_1)+\omr_i(k_2)} +\frac{n(\omr_i(k_1))-n(\omr_i(k_2))}{\omr_i(k_2)-\omr_i(k_1)} \right\}, \nonumber \\
(A,A)_{yy}(q)&=&\half\sum_{\v{k}}\frac{|V_3(qy,k_1 x, k_2 y)|^2}{\omr_x(k_1) \omr_y(k_2)} \left \{ \frac{1+n(\omr_x(k_1))+n(\omr_y(k_2))}{\omr_x(k_1)+\omr_y(k_2)} +\frac{n(\omr_x(k_1))-n(\omr_y(k_2))}{\omr_y(k_2)-\omr_x(k_1)}.
\right \}. \nonumber \\
\end{eqnarray}
In order to analyze the low wave-vector $(q\to 0)$ behavior of $D(q)$, we simplify the equations by noting that
\begin{eqnarray}\label{approxVN}
 V_3(qi,1,2) \propto \Delta(q+k_1+k_2) \sin(q)\sin(k_1)\sin(k_2)&\xrightarrow{q \to 0}& \Delta(q+k_1+k_2) q \sin^2(k_1),\nonumber\\
\frac{n(\omr_{\alpha}(k_1))-n(\omr_{\alpha}(-k1-q))}{\omr_{\alpha}(-k_1-q)-\omr_{\alpha}(k_1)} &\xrightarrow{q\to 0}& -\frac{\partial}{\partial \omr}n(\omr_{\alpha}(k_1)).
\end{eqnarray}
We first consider $(A,A)_{xx}(q)$, whose expression contains two terms corresponding to scattering by longitudinal ($i=x$) and transverse ($i=y$) modes respectively. We may first suppose that the renormalized frequencies have the same $q$-dependence as the bare frequencies, $\omr(q) \propto \om(q)$. Then the use of Eq.~(\ref{approxVN}) in Eq.~(\ref{DEqDiagonal}) shows that, as $q\to 0$,  the longitudinal contribution to $(A,A)_{xx}(q)$ behaves as $q^2$ whereas the transverse part gives rise to a divergence basically due to $q^2$-dependence of the transverse frequencies in the denominator. Similarly the expression for $(A,A)_{yy}(q)$ also involves transverse mode frequencies in the denominator, which again give rise to a singular dependence on $q$. 

If on the other hand we assume $\omr_{\alpha}(q\to0)\propto q$, all the integrals converge and we get $D_{\alpha \alpha}(q\to0) \propto q^2$. Thus we have a self-consistent solution for $D_{\alpha \alpha}(q)$ in the $q\to 0$ limit, which is also diagonal. This is exactly what is seen in the MD simulations of the classical system with the Wang-Li interaction~\cite{WangLi}. From this analysis it is clear that the linear solution obtained here is independent of the form of bare frequencies. Thus the linear $q$-dependence of the frequencies in one dimension seems to be a generic result. For further analysis we take a specific form for renormalized phonon frequencies, namely,
\begin{equation}
 \label{renormalisedFrequency}
\omr_i(q)\equiv \omr_i \omr(q), \quad \omr(q)=2 |\sin(q/2)|, \quad \omr_y=\eta \omr_x,
\end{equation}
which agrees with the self-consistent solution in the $q\to 0$ limit. 

\section{Relaxation rate of renormalized phonons}
We now study an effective Hamiltonian given by Eq.~(\ref{HcubicInNumberOp}) and Eq.~(\ref{aDefinition}) in which $\om(q)$ is replaced by $\omr(q)$. The second-order relaxation rate for the renormalized modes are as before,
\begin{eqnarray}\label{GammxRenormalized}
\Gamma_x^{(2)}(q)&=&\frac{1}{4} (1-e^{-\beta \omr_x(q)})\int dk \sum_{i s_1 s_2} |\vt_3(-q \, x,k\,  i,q-k\,  i)|^2  n_{s_1}(\omr_i(k)) n_{s_2}(\omr_i(q-k)) \nonumber\\ 
&&\times \delta(\omr_x(q)-s_1\omr_i(k)-s_2\omr_i(q-k))\\
\label{GammyRenormalized}
 \Gamma^{(2)}_y(q)&=&\half \left( 1-e^{-\beta \omr_y(q)} \right ) \int dk \sum_{s_1 s_2} |\vt_3(-q\, y,k\, x,q-k\, y)|^2  n_{s_1}(\omr_x(k)) n_{s_2}(\omr_y(q-k)) \nonumber\\ 
&&\times \delta(\omr_y(q)-s_1\omr_x(k)-s_2\omr_y(q-k)),\\
&&\vt_3(k_1 i_1,k_2 i_2,k_3 i_3)=V(k_1 i_1,k_2 i_2,k_3 i_3)/\sqrt{\omr_{i_1}(k_1) \omr_{i_2}(k_2) \omr_{i_3}(k_3)}
\end{eqnarray}
 First we estimate the low-$q$ form of $\Gamma^{(2)}_y(q)$. This proceeds exactly as the analysis of  Eq.~(\ref{gamm2BarePhonon}). We have 
\begin{eqnarray}
 \Gamma^{(2)}_y(q)=\frac{1}{2} (1-e^{-\beta \omr_y(q)})\sum_{s_1,s_2,k^*}  \frac{|\vt_3(-q y,k^*\, x,q-k^*\, y)|^2}{|J(q,k^*,s_1,s_2)|}  n_{s_1}(\omr_x(k^*)) n_{s_2}(\omr_y(q-k^*)),
\end{eqnarray}
where $k^*$ are the solutions of $f\equiv \omr_y(q)-s_1\omr_x(k)-s_2\omr_y(q-k)=0$ and the Jacobian is given by $J=\partial f/\partial k|_{k=k^*}$. Since both $\omr_{x,y}(q\to 0) \sim q$ the factors of $q$ in the denominator that arise due to number factors are canceled by the $|\vt|^2$ term. Therefore if the Jacobian $J(k^*)$ does not go to zero at some $k^*$, we get  $\Gamma^{(2)}_y(q \to 0) \propto \omr_y^2(q)$.
 
\begin{figure}[t]
\begin{center}
\includegraphics  [width=15cm]{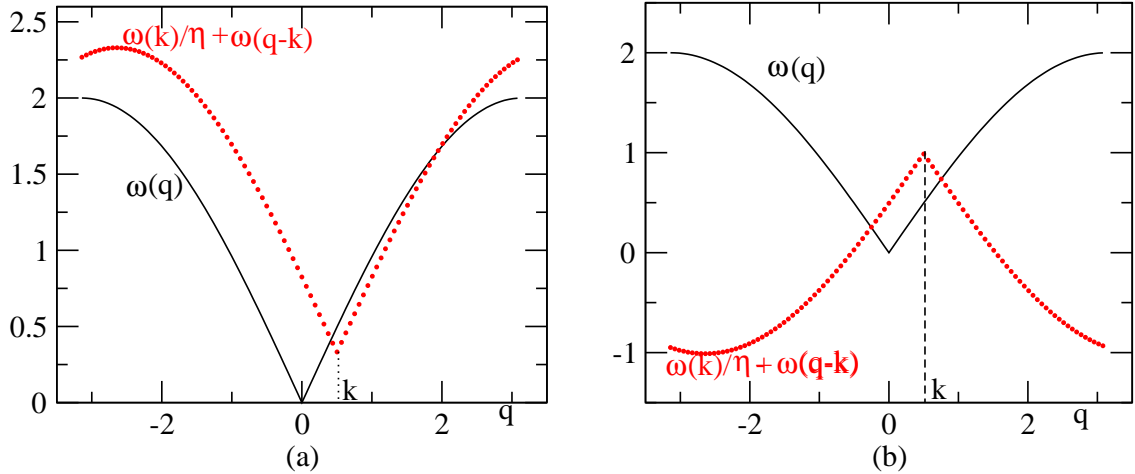}
\caption{Figures demonstrate how the energy conservation condition in Eq.~(\ref{GammyRenormalized}) can be satisfied. The figure on the left shows $\omr(q)$ and $\omr(k)/\eta+\omr(q-k)$ in the same graph for $k=0.5, \eta=1.5$ and the figure on the right shows $\omr(q)$ and $\omr(k)/\eta-\omr(q-k)$ for $k=0.5, \eta=0.5$. Dotted vertical lines mark the values of $k$ which is the solution for $q$ values corresponding to the intersections of the two graphs.}
\label{FigomegaEquation}
\end{center}
\end{figure}

There are four cases for the energy-momentum conservation conditions due to four possible values of $s_1$ and $s_2$. These have to be analyzed separately. We choose the Brillouin zone $[0,2\pi]$. For $s_1=+=s_2$, the energy conservation condition becomes $\omr(q)-\omr(k)/\eta-\omr(q-k)=0$. When $\eta=1$ there are two solutions $k=0$ and $k=q$. But these correspond to no scattering and have to be excluded from the perturbation terms. Thus as noted earlier the relaxation rate within this approximation is zero~\cite{SanthoshDeepakOddPot}. When $\eta >1$ nontrivial solutions are possible. Fig.~\ref{FigomegaEquation}(a) shows plots of $\omr(q)$ and  $\omr(k)/\eta+\omr(q-k)$ with respect to $q$ for a representative set of values for $k$ and $\eta$. This figure shows that, when $\eta>1$, $k$ is a solution for the energy conservation condition $\omr(q)=\omr(k)/\eta+\omr(q-k)$ when the value of $q$ corresponds to one of the intersections of the two graphs.  For small-$q$ we get a solution $k\sim q$ with a finite value for the Jacobian of the argument of the $\delta$-function. Therefore we see that the contribution to $\Gamma^{(2)}_y(q\to 0)$ from this case is proportional to $q^2$ if $\eta>1$ and zero otherwise.

For $s_1=+,s_2=-$, the $\delta$-function condition is $\omr(q)=\omr(k)/\eta-\omr(q-k)$. Fig.~\ref{FigomegaEquation}(b) shows LHS and RHS of this condition on the same graph. We see that for $\eta < 1$ we have a nontrivial solution and in the small-$q$ limit we find a solution $k\sim q$ where the Jacobian also is finite. Therefore, for this case, the contribution to $\Gamma^{(2)}_y(q\to 0)$ is proportional to $q^2$ for $\eta<1$ and zero otherwise. 

Next we take $s_1=-,s_2=+$. The energy conservation is given by $\omr(q)=\omr(q-k)-\omr(k)/\eta$. We have a nontrivial solution for  $\eta >1$ which for $q\to0$ is $k\sim q$. Again we find the contribution to $\Gamma^{(2)}_y(q\to 0)$ is like $q^2$ for $\eta>1$ and zero otherwise. The fourth case $s_1=s_2$ has only trivial solutions. Therefore, we have 
\begin{equation}\label{sopsol}
 \Gamma^{(2)}_y(q\to 0) \sim q^2 \quad \mbox{if } \eta \neq 1.
\end{equation}

Now we consider the relaxation rate $\Gamma^{(2)}_x$. It has contributions coming from both transverse and longitudinal modes corresponding to $i=y$ and $i=x$  in Eq.(~\ref{GammxRenormalized}). The $i=y$ term again involves interaction between two $y$-phonons and one $x$-phonon as in the case for $\Gamma^{(2)}_y$ and its low wave-vetor behavior can be calculated in a similar way. As in the calculation of $\Gamma^{(2)}_y$, for $\eta\neq 1$, there are nontrivial solutions satisfying the $\delta$-function except for $s_1=s_2=-1$. Thus the corresponding contribution to $\Gamma^{(2)}_x(q\to 0)$ is proportional to $q^2$. The $i=x$ term involves three $x$-phonons and the contribution to the relaxation rate is zero because the $\delta$-function condition $\sin(q/2)-s_1 \sin(k/2)-s_2\sin((q-k)/2)=0$ has no nontrivial solution. Therefore 
\begin{equation}
 \Gamma^{(2)}_x(q\to 0) \sim q^2 \quad \mbox{if } \eta \neq 1.
\end{equation}

Thus, the renormalized phonon modes at low wave-vectors are slow decaying and well-defined to perform the perturbation analysis.

\section{Self-consistent relaxation rate}
The phonon relaxation rates of a chain with cubic nonlinearity but with only the longitudinal vibrations was studied earlier by us ~\cite{SanthoshDeepakOddPot}. For this case, as mentioned in the above paragraph, the relaxation rate is zero in the second order as the two conservation conditions for energy and wave-vector cannot be satisfied simultaneously. To overcome this difficulty we performed a self-consistent approximation which consists in replacing bare phonon Green's functions in the second-order diagram (Fig.\ref{selfEnergyDiagrams}(a)) by full Green's functions. This relaxes the strict energy conservation condition and allows for nonzero solutions for the relaxation rate through a solution of an integral equation. Our aim in this section is to perform a similar analysis for the present case. For this purpose, one uses the spectral representation of the Green's function,
\begin{equation}
 D_i(q,\tau)=-\sum_{s\in\{+,-\}}\int \frac{d\epsilon}{2\pi} B_{i}(k,\epsilon) n_s(\epsilon)e^{-s\epsilon \tau},
\end{equation}
with the spectral function is given by
\begin{equation}
 B_i(k,\ep)=\frac{2\Gamma_{i}(k,\ep)}{(\ep-\tilde{\omr}_{i}(k,\ep))^2+\Gamma^2_{i}(k,\ep)},
\end{equation}
where $\tomr_{i}(k,\ep)=\omr_i(k)-\Sigma_i^R(q,\ep)$ and $\Sigma_i^R(q,\ep)=\Re\,  \Sigma_i(q,i\omr_n\to\ep+i0^+)$. If we now replace $D^{(0)}(q,\tau)$ 
by $D_i(q,\tau)$ in Eq.~(\ref{sigma}), the corresponding equations for relaxation rate, Eqs.~(\ref{GammxRenormalized}) and (\ref{GammyRenormalized}), are replaced by a set of self-consistent equations for $\Gamma_x$ and $\Gamma_y$.  We get 
\begin{eqnarray}\label{selfconstFirst}
 \Gamma_x(q)&=&\frac{1}{2N}(1-e^{-\beta \omr_x(q)}) \sum_{\v{k},\v{s},i} |V_3(-qx,k_1i,k_2i) |^2 \int \frac{d\ep}{4\pi} B_i(s_1k_1,\ep)n_{s_1}(\ep)\nonumber \\
&&\times B_i(s_2k_2,s_2\omr_x(q)-s_1s_2\ep)n_{s_2}(s_2\omr_x(q)-s_1s_2\ep),\nonumber\\
\Gamma_y(q)&=&\frac{1}{N} (1-e^{-\beta \omr_y(q)}) \sum_{\v{k},\v{s}} |V_3(-qy,k_1x,k_2y) |^2 \int \frac{d\ep}{4\pi} B_x(s_1k_1,\ep)n_{s_1}(\ep)\nonumber \\
&&\times B_y(s_2k_2,s_2\omr_y(q)-s_1s_2\ep)n_{s_2}(s_2\omr_y(q)-s_1s_2\ep).
\end{eqnarray}
We follow Ref~\cite{SanthoshDeepakOddPot} in the analysis of these equations. We note that for small $q$, the relaxation rate $\Gamma_i(q)$ is very small and the spectral function $B_i(q,\ep)$ can be approximated by a narrow peak at $\ep=\tomr_i(q,\omr_i(q))=:\tomr_i(q)$. Also, $\tomr_i(q \to 0) \propto \omr_i(q)$.  Further we may replace $B_i(p,\ep) f(\ep)\approx B_i(p,\ep)f(\tomr_i(p))$ where $f(\ep)$ is a smooth function of $\ep$. We use this approximation for the number factors $n(\ep)$ in Eq.~(\ref{selfconstFirst}). We also replace $\tomr_i(q,\ep)\approx \tomr_i(q)$ and $\Gamma_i(q,\ep)\approx \Gamma_i(q)$ in the expression for $B_i(q,\ep)$ to be used in Eq.~(\ref{selfconstFirst}). After these approximations, only the spectral functions depend on $\ep$ in Eq.~(\ref{selfconstFirst}) and the integral can be performed using
\begin{eqnarray}\label{IntFormula}
 \int \frac{d\ep}{\pi} \frac{\Gamma_1}{(\ep-a)^2+\Gamma_1^2}\frac{\Gamma_2}{(\ep-b)^2+\Gamma_2^2}=\frac{(\Gamma_1+\Gamma_2) [(a-b)^2+(\Gamma_1-\Gamma_2)^2]}{((a-b)^2+\Gamma_1^2+\Gamma_2^2)^2-4\Gamma_1^2\Gamma_2^2}.
\end{eqnarray}
We note that $\tomr_i(q)=\tomr_i(-q)$ because it depends on $q$ through $\omr(q)$ which is even in $q$ and through $V_3$ factors which enters the perturbation terms as even function of $q$ only. Then, $\Gamma_i(q)=\Gamma_i(-q)$ and  $n(q)\equiv n(\tomr(q))=n(-q)$. We can write $\Gamma_x(q)=\Gamma_x^A(q)+\Gamma_x^B(q)$, with
\begin{eqnarray}
\label{selfConstGxA}
 \Gamma_x^A(q)&=&\frac{1}{2N}(1-e^{-\beta \omr_x(q)})\sum_{\v{k} i}n(\omr_i(k))\big \{ n(\omr_i(q+k))|V_3(-qx,-k i,-q-k\,i))|^2 \nonumber\\
&&\times S_{++}\left(qx,k i,q+k\ i\right) +e^{\beta \omr_x(q)} n(\omr_i(q-k))|V_3(-qx,k i,q-k\,i)|^2 \nonumber\\ &&\times  S_{--}\left(qx,k i,q-k\,i\right) \big\}, \\
\label{selfConstGxB}
\Gamma_x^B(q)&=&\frac{1}{N}(1-e^{-\beta \omr_x(q)})\sum_{\v{k} i}|V_3(-qx,-ki,q+k\,i)|^2  n(\omr_i(k))[1+n(\omr_i(q+k))] \nonumber \\
&&\times S_{+-}(qx,k i,q+k\,i),
\end{eqnarray}
where 
\begin{eqnarray}\label{Sfactor}
 S_{s_2s_3}(k_1 i_1,k_2 i_2,k_3 i_3)=\frac{(\Gamma_{i_2}(k_2)+\Gamma_{i_3}(k_3))[ \chi^2_{s_1s_2}(k_1i_1,k_2i_2,k_3i_3)+(\Gamma_{i_2}(k_2)-\Gamma_{i_3}(k_3))^2 ]}{[\chi^2_{s_1s_2}(k_1i_1,k_2i_2,k_3i_3) +\Gamma^2_{i_2}(k_2)+\Gamma^2_{i_3}(k_3)]^2-4\Gamma^2_{i_2}(k_2)\Gamma^2_{i_3}(k_3)}.
\end{eqnarray}
Here we have introduced the definition: $\chi_{s_1s_2}(k_1i_1,k_2i_2,k_3i_3)=\omr_{i_1}(k_1)+s_2\omr_{i_2}(k_2) + s_3\omr_{i_3}(k_3)$. Similarly, for the transverse modes we get
\begin{eqnarray}\label{selfConstGy}
 \Gamma_y(q)&=&\frac{1}{N}(1-e^{-\beta\omr_y(q)}) \sum_k \big\{ n(\omr_x(k))  n(\omr_y(q+k)) |V_3(-qy,-kx,q+k\,y)|^2 S_{++}(qy,kx,q+k\,y)\nonumber\\
&& + e^{\beta \omr_y(q)}n(\omr_x(k)) n(\omr_y(q-k))|V_3(-qy,kx,q-k\,y)|^2 S_{--}(qy,kx,q-k\,y)  \nonumber\\
&& + n(\omr_x(k))[1+n(\omr_y(q+k))]|V_3(-qy,-kx,q+k\,y)|^2 S_{+-}(qy,kx,q+k\,y) \nonumber \\
&& + [1+n(\omr_x(k))]n(\omr_y(q-k)) |V_3(-qy,kx,q-k\,y)|^2 S_{-+}(qy,kx,q-k\,y)
\big\}.
\end{eqnarray}

Equations~(\ref{selfConstGxA}), (\ref{selfConstGxB}) and (\ref{selfConstGy}) have to solved self-consistently to obtain the small wave-vector behavior of the relaxation rates. We show that these equations have a solution
\begin{equation}\label{GammaSelfConst}
\Gamma_x(q\to 0)\sim q^{3/2}, \quad \quad \Gamma_y(q\to 0)\sim q^2. 
\end{equation}
The key feature that enables the solution of these equations is that for small $q$, each of the equations can be written in the following schematic form
\begin{equation}\label{Fkq}
\Gamma_i(q\to 0) \propto \beta \omega_i^2(q) \int dk ~ F(k,q).
\end{equation}
Here one factor of $\omega_i(q)$ comes from the prefactor, which is basically a detailed balance factor and the other from the factor of $|V_3|^2$. The integrand $F(k,q)$ is not known as it as it depends on $\Gamma_i(k)$'s. Our strategy is to examine these integrands in Eqs.~(\ref{selfConstGxA}), (\ref{selfConstGxB}), (\ref{selfConstGy}) at $q=0$, by first assuming that $\Gamma_i(k) \propto k^2$. If the integral turns out to be finite, clearly $\Gamma_i(q) \propto q^2$ is a self-consistent solution. If on the other hand the integral diverges, it presumably behaves for small $q$ as $q^{-x}$. So we assume that $\Gamma(k)\sim k^{\delta}$ with $\delta < 2$, and examine the integral for small $q$. A solution is achieved if a self-consistent value for $\delta$ can be obtained.
In examining the singularities of the integrand in Eq.~(\ref{Fkq}), a further simplification comes from noting that the Bose factors $n(k)$ give rise to factors $1/\omega_i(k)$ for small frequencies, but these are canceled by the similar factors in interaction term $|V|^2$. So all the singular behavior of the integrands come from the $S$-factors introduced in Eq.~(\ref{Sfactor}).

First, let us consider Eq.~(\ref{selfConstGy}), the equation for $\Gamma_y(q)$. At $q=0$, the $S$-factors could give rise to singularities at $k=0$ if we set $\Gamma(k) \propto k^2$.  For $\eta \neq 1$, in all the four terms in Eq.~\ref{selfConstGy}, $\chi(0,kx,ky)\sim k$ for small $k$; for example, in the third term, $\chi_{+-}(0,kx,ky)=2(1-\eta)\sin(k/2)$. Using Eq.~(\ref{GammaSelfConst}) one finds that $S_{s_1s_2}(0,kx,ky)\sim k^{2} k^2/k^4=k^{0}$ for small $k$. Since the integral is finite at $q=0$, $\Gamma_y(q\to 0)\sim q^2$. 
We next consider $\Gamma_x(q)$ which has been split into two terms $\Gamma_x^A(q)$ and $\Gamma_x^B(q)$ in Eqs. (\ref{selfConstGxA}) and (\ref{selfConstGxB}). The analysis of $\Gamma_x^A(q)$ proceeds in a similar way as $\chi(0,ki,ki)\sim k$ as $k\to 0$  in both the terms of Eq.~(\ref{selfConstGxA}) and we get $\Gamma_x^A(q)\to 0)\sim q^2$. 

Something different happens in Eq.~(\ref{selfConstGxB}) for $\Gamma_x^B(q)$. It has again two terms for $i=x,y$. The term $i=x$ involves only the longitudinal mode, and this situation is identical to the FPU-$\alpha$ problem discussed in Ref.~\cite{SanthoshDeepakOddPot}. We note that $\chi_{+-}(q,kx,q+k\,x)\sim q[1-\cos(k/2)]\propto q k^2$ as $q,k \to 0$. Using Eq.~(\ref{GammaSelfConst}), for small $q$ and $k$, we get $S_{+-}(qx,kx,q+k\,x)\propto [\Gamma_x(q)+\Gamma_x(q+k)]^{-1}$. At $q=0$, the corresponding integral is divergent. But the small $q$ divergence of the integral can be obtained by noting that the major contribution to the integral comes from small $k$ and small $q+k$ region.  Denoting the corresponding contribution to $\Gamma_x^B$ by $\Gamma_1$, we get, following~\cite{SanthoshDeepakOddPot},
\begin{eqnarray}\label{Gamma1}
 \Gamma_1(q)\propto q^2\int_0^{2\pi}\frac{dk}{\Gamma_1(q+k)+\Gamma_1(k)}.
\end{eqnarray}
As is easily checked, Eq.~(\ref{Gamma1}) is solved by taking $\Gamma_1(k)\propto k^{3/2}$. 
For the $i=y$ term $\chi_{+-}(qx,ky,q+k\,y)\sim q[1-\eta \cos(k/2)]$ for small $q$ and $k$.  The leading order behavior of $S$ in this limit  is obtained using $\Gamma_x(q\to 0)\sim q^2$. We see that $S_{+-}(qx,ky,ky)$ is not singular at $k=0$ for finite $q$,  $S_{+-}(qx,0 y,q y) \sim q^2 q^2/q^4 \sim q^0$. Thus the limit $q \to 0$ of this sequence is finite and the corresponding contribution to $\Gamma_x^B(q\to0)$ is proportional to $q^2$. We can now go back to the earlier equations and check if by setting $\Gamma_x(k)\propto k^{3/2}$. the earlier results for $\Gamma_y(q)$ and $\Gamma_x^A(q)$ are affected. The examination of the integrals shows that they remain finite. Therefore to the leading order, $\Gamma_x(q\to0)\sim q^{3/2} + O(q^2)$ and  $\Gamma_y(q\to 0)\sim q^2$ establishing the solution given by Eq.~(\ref{GammaSelfConst}).

\section{Thermal conductivity}
In this Section, we calculate the thermal conductivity of the system using the Kubo-Green formula. Defining the thermal Green's function for the total current operator $J$ as,
\begin{equation}\label{defW}
 W(\tau)=-\frac{1}{N} \left \langle T_{\tau} \left[ J(\tau) J(0)\right] \right \rangle,
\end{equation}
the  Kubo-Green formula for thermal conductivity can be written as ~\cite{MahanBook}
\begin{equation}\label{KuboFormula}
 \kappa=-\frac{1}{T}\lim_{\omega \to 0}\lim_{N\to \infty}\Im \frac{W(i\omega_n \to\omega+i\delta)}{\omega}
\end{equation}
where $\omega_n$ are the Matsubara frequencies. The current operator, up to leading order, is given by
\begin{equation}
 J=-\sum_{k,\alpha}  j_{\alpha}(k) \d{b}_{\alpha}(k)b_{\alpha}(k)
\end{equation}
where $j_{\alpha}(k)=v_{\alpha}(k) \om_{\alpha}(k)$, $v_{\alpha}(k)=(\partial/\partial k) \omr_{\alpha}(k)$.  Further, we neglect the cross-correlation between the x- and y- components of the current. Under the approximation used in ~\cite{SanthoshDeepakFPUbeta} that  vertex corrections to the two particle Green's function Eq.~(\ref{defW}) has the same wave-vector dependence as $j(k)$ for small $k$, one arrives at the kinetic theory formula for conductivity,
\begin{equation}\label{conductivityKineticTheoryFormula}
 \kappa=\sum_{\alpha}\int \frac{dq}{2\pi} v_{\alpha}^2(q) \tau_{\alpha}(q) C_{\alpha}(q) ,
\end{equation}
where $\tau_{\alpha}(q)=\Gamma_{\alpha}^{-1}(q)$ is the relaxation time of single particle Green's function and $C_{\alpha}(q)=\frac{1}{T}(-\frac{d}{d\omr}n(\omr))_{\omr=\omr_{\alpha}(q)}\omr_{\alpha}^2(q)$ is proportional to the mode specific heat. We note that $C(q)$ is well behaved for small $q$ and any possible divergence can come from $\tau(q)$ only. 

The formula given by Eq.~(\ref{conductivityKineticTheoryFormula}) is for an infinite chain and this quantity is divergent because $\tau_{\alpha}(q)$ has a non integrable singularity at $q=0$. To obtain the conductivity $\kappa(N)$ for large but finite size chain of $N$ particles, we follow the prescription used by other workers in the area~\cite{LepriLiviPolitiPhysRep}. In this procedure one rewrites Eq.~(\ref{conductivityKineticTheoryFormula}) as a time integral in the following way.
\begin{equation}\label{KuboFormulaNonEq}
 \kappa=\sum_{\alpha}\int_0^{\infty} dt K_{\alpha}(t),\quad K_{\alpha}(t)=\int \frac{dq}{2\pi} v_{\alpha}^2(q) C_{\alpha}(q) e^{-\Gamma_{\alpha}(q)t}
\end{equation}
Since $\Gamma_{\alpha}(q\to0)\sim q^{\delta_{\alpha}}$, we get $K_{\alpha}(t\to\infty) \sim t^{-1/\delta_{\alpha}}$. One argues that when the boundaries of the system are connected to thermal baths, the correlation $K(t)$ dies after a time scale of order $N$ due to interaction with the baths, i.e. the typical time it takes for the modes to travel across the chain. Using in the time integral in Eq.~(\ref{KuboFormulaNonEq}) such a cutoff yields
\begin{equation}
 \kappa(N) \sim O(N^{1-1/{\delta_x}})+O(N^{1-1/{\delta_y}})\sim O(N^{1/3})+O(N^{1/2})\sim N^{1/2}.
\end{equation}
It is  to be noticed that the leading order divergence arise from the term corresponding to the $y$-modes. If the conduction was normal, conductivity would be finite and the contribution from these modes would be negligible compared to the contribution from the $x$-modes when the strength of the interaction giving rise to transverse oscillations, $K_{\theta}$, is small. But here the conductivity itself is a divergent quantity and its divergence is determined by the slower relaxation of the transverse modes compared to the longitudinal modes. 

\section{Summary}
We have studied thermal transport in a chain of coupled particles in which particles can have longitudinal as well as transverse vibrations. Particles interact via anharmonic potentials proposed by Wang and Li~\cite{WangLi}. For this model, the mode frequencies in the harmonic approximation have the following dispersions at small wave-vectors: for the longitudinal mode $\Omega_x(q) \propto q$; for the transverse mode $\Omega_y(q) \propto q^2$, where $q$ denotes the wave-vector. We first show that due to quadratic dispersion of the transverse mode, the relaxation rates for both the modes become too large to leave them as meaningful quasi-particles. So we calculate their renormalized frequencies taking into account the anharmonic interactions in a self-consistent approximation. We find that this procedure yields the linear dispersion for both kinds of modes irrespective of their bare forms. We then perform a calculation of the relaxation rates for the renormalized phonon in a self-consistent approximation based on the second-order perturbation diagrams. This calculation yields that for the longitudinal phonons, the relaxation rate $\Gamma_x(q) \propto q^{3/2}$, while for the transverse mode $\Gamma_y(q) \propto q^{2}$. We then use these results to calculate the thermal conductivity $\kappa(N)$ of a chain of $N$ particles. The finite size is incorporated by using a cutoff of the order of $N$ in the time integral over current-current correlation function that occurs in the Kubo formula for the conductivity. This yields a contribution to $\kappa(N)$ from longitudinal phonons to be $O(N^{1/3})$, that from transverse phonons to be $O(N^{1/2})$. The latter contribution dominates giving $\kappa(N) \propto N^{1/2}$ for large N.

\end{document}